\documentclass[aps,prl,,showpacs,twocolumn,groupedaddress]{revtex4}

\usepackage{amsmath,amssymb}
\usepackage{graphicx}
\usepackage{psfrag}

\newcommand{\lsbar}{\langle l_s\rangle}

\begin{document}

\preprint{LMU-ASC 12/05}

\title{Stiff Polymers, Foams and Fiber Networks}
\author{Claus Heussinger}
\author{Erwin Frey}

\affiliation{Arnold Sommerfeld Center for Theoretical Physics and
  CeNS, Department of Physics, Ludwig-Maximilians-Universit\"at
  M\"unchen, Theresienstrasse 37, D-80333 M\"unchen, Germany}

\affiliation{Hahn-Meitner-Institut, Glienicker Strasse 100, D-14109
  Berlin, Germany}

\begin{abstract}
  We study the elasticity of fibrous materials composed of generalized
  stiff polymers. It is shown that in contrast to cellular foam-like
  structures affine strain fields are generically unstable. Instead, a
  subtle interplay between the architecture of the network and the
  elastic properties of its building blocks leads to intriguing
  mechanical properties with intermediate asymptotic scaling regimes.
  We present exhaustive numerical studies based on a finite element
  method complemented by scaling arguments.
\end{abstract}

 \date{\today}

\maketitle

\begin{figure}[htb!]
\begin{center}
  \includegraphics{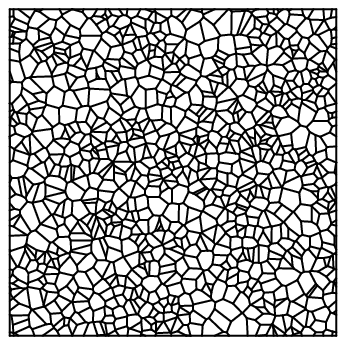} \hspace{1cm}
  \includegraphics{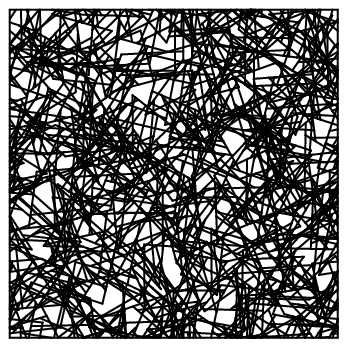}
\end{center}
\caption{\label{fig:vorNet} Illustration of the different
  architecture of (a) cellular and (b) fibrous materials. The foam in
  (a) is constructed by a Voronoi tessellation from the centers of the
  fibers in (b).}
\end{figure}

Cellular and fibrous materials are ubiquitous in nature and in many
areas of technology. Examples range from solid or liquid foams over
wood and bone to the protein fiber network of
cells~\cite{wea01,gib99,alb94}. On a mesoscopic level these materials
are comprised of struts and membranes with anisotropic elastic
properties.  The systems differ widely in architecture. One finds
patterns which are as regular as a honeycomb, as sophisticated as the
particular design of a dragonfly's wing or simply random~\cite{tho61}.
The manifold combinations of architecture and elastic properties of
the building blocks allows for a rich spectrum of macroscopic elastic
responses.
For cellular structures (see the foam-like structure in
Fig.~\ref{fig:vorNet}a) macroscopic elasticity can already be
understood by considering the response of a single
cell~\cite{gib99,kra94}.  In these systems local stresses acting on an
individual cell are the same as those applied on the macroscopic
scale. In other words, the local deformation $\delta$ of a cell with
linear dimension $l_s$ follows the macroscopic strain $\gamma$ in an
\emph{ affine way} such that it scales as $\delta \propto \gamma l_s$.
This affinity of the deformations seems to be a reasonable
approximation even for systems with some randomness in the size and
the type of their unit cells \cite{faz02,sil95}.  Since in affine
models there can be no cooperativity between the elastic responses of
individual cells, the effect of the assembled structure can simply be
predicted by counting the numbers of cells.
Fibrous networks, on the other hand, are dramatically different
already in their morphology as can be inferred from
Fig.~\ref{fig:vorNet}. The presence of fibers introduces the
additional mesoscopic scale of the fiber length $l$ and generates a
broad distribution of pore sizes that, in contrast to foams, has a
non-vanishing weight even for the smallest cells~\cite{kal60}. This
difference in architecture crucially affects the mechanical
properties. Recently a non-affine regime has been
identified~\cite{frey97} and characterized~\cite{wil03,hea03a} in
two-dimensional networks of classical beams (``Mikado model'')commonly
used to model the mechanical properties of paper sheets
\cite{ast00,ast94,lat01,hey96}.  The non-affinity of the deformation
field necessarily implies that in these networks cooperativity effects
play an important role.

In this Letter we will address the question of how such non-trivial
network geometry generates correlation-induced linear elastic behavior
while cellular foam-like structures can generally be described by
mean-field like approaches.  By systematically tuning the force
response properties of the individual elements we will be able to
identify a new length-scale below which correlations drive the system
away from the state of affine deformations. We will, moreover,
describe the mechanism that generates this length and calculate the
resulting power law behavior of the elastic modulus by a scaling
argument.

The model is defined as follows.  $N$ anisotropic elastic elements,
geometrically represented by straight lines of length $l$, are placed
on a plane of area $A=L^2$ such that both position and orientation of
the elements are uniformly random distributed. The length of the
segments, i.e. the distance $l_s$ between any two neighboring
intersections, follows an exponential distribution \cite{kal60}
\begin{equation}\label{eq:segDist}
P(l_s)=\lsbar^{-1} e^{-l/\lsbar}\,,
\end{equation}
with a mean value that is given in terms of the density $\rho=Nl/A$ as
$\lsbar= \pi / 2\rho $.  At any intersection a permanent crosslink
with zero extensibility is generated. This constrains the relative
translational motion of the two filaments, while leaving the
rotational degrees of freedom independent (free hinges).
Next to the architecture of the network we have to specify the linear
elastic properties of the constituents. Previous studies
\cite{wil03,hea03a} have considered classical beams of radius $r$ and
bending stiffness $\kappa$.  Loaded along their axis (``stretching'')
such slender rods have a rather high stiffness $k_s=4\kappa/l_sr^2$,
while they are much softer with respect to transverse deformations
$k_\perp(l_s)=3\kappa/l_s^3$ (``bending'').  Here, we consider elastic
elements where in addition to the mechanical stiffness of classical
beams a more general stretching coefficient
\begin{equation}\label{eq:k_entr}
  k_{\parallel} (l_s) =  
  6 \, \kappa \, \frac{l_p^{\alpha}}{l_s^{3+\alpha}}\,,
\end{equation} 
is introduced. This may result from thermal fluctuations of the
filament immersed in a heat bath of solvent molecules. The prefactor
is chosen such that $k_{\parallel}$ for $\alpha = 1$ reduces to the
longitudinal entropic elasticity of a stiff polymer described by the
wormlike chain model grafted at one end~\cite{kro96}.  In this case
the material length $l_p$ is called the persistence length of the
polymer and quantifies the ratio of bending to thermal energy
$l_p=\kappa/k_BT$.  The phenomenological exponent $\alpha$ allows us
to extend our discussion to the broad class of systems for which
$k_\parallel$ is a monomial (with units energy per area) involving one
additional material length $l_p$.  Having two longitudinal deformation
modes the effective stretching stiffness is equivalent to a serial
connection of the ``springs'' $k_s$ and $k_\parallel$.  Setting $l_p=c
r$, we can write $k_s \propto k_\parallel(\alpha=-2)$.  The constant
$c$ is a material property of the specific polymer and has been chosen
as $c=1.5\cdot 10^4$, which roughly corresponds to the biopolymer
F-actin. The precise value, however, is irrelevant what regards the
thermal response $k_\parallel$ and only specifies the location of the
cross-over to $k_s$.
The description of a thermally fluctuating network in terms of force
constants $k_\parallel$, $k_s$ and $k_\perp$ is in the spirit of a
Born-Oppenheimer approximation that neglects the fluctuations of the
``slow variables'', the cross-link positions, while assuming the
``fast'' polymer degrees of freedom to be equilibrated. By
minimization of the internal energy with respect to the slow
parameters, the modulus for a given macroscopic deformation can be
calculated. In practice, we apply a shear strain of $\gamma=0.01$ and
thus determine the shear modulus $G$ using periodic boundary
conditions on all four sides of the simulation box. The minimization
procedure is performed with the commercially available finite element
solver MSC.MARC.

\begin{figure}
  \psfrag{G/kr3}{\small$G/\kappa\rho^3$}
  \psfrag{rlp}{\small$\rho l_p$}
  \psfrag{y45}{\small$l_p^{0.46}$}
  \psfrag{y9}{\small$ l_p^{0.9}$} \psfrag{y1}{\small$l_p^{1}$}
\begin{center}  
  \includegraphics[width=0.9\columnwidth]{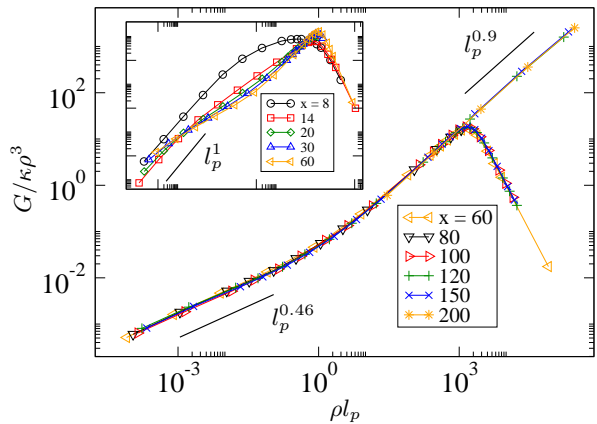}
\end{center}
\caption{Scaling function $g$ as a function of $\rho l_p$ for
  various $x=\rho l\ge60$; Inset: The same for $x \le 60$. }
  \label{fig:modulusScaled}
\end{figure}

To set up a reference frame, it is instructive to discuss a mean-field
approach which assumes that correlations between neighboring segments
are absent. As outlined in the introduction this corresponds to a
foam-like behavior, where the response is fully described by the
properties of an average segment of length $\lsbar\propto\rho^{-1}$.
Marking the force constants of this segment by an overbar, we can
express them in the form (neglecting numerical prefactors)
$\bar{k}_\perp\simeq\kappa\rho^3$,
$\bar{k}_\parallel\simeq\bar{k}_\perp(\rho l_p)^\alpha$ and
$\bar{k}_s\simeq\bar{k}_\perp(\rho r)^{-2}$, respectively.  The
corresponding deformation modes will act as springs connected in
series~\cite{kra94} such that the modulus takes the form
\begin{equation}\label{eq:modFoam}
G_{\rm foam}^{-1} = 
a\bar{k}_\parallel^{-1}+b\bar{k}_\perp^{-1}+c\bar{k}_s^{-1}\,.
\end{equation}
The network will thus show a crossover from thermal stretching to
bending at $l_p\approx\lsbar$ and to mechanical stretching at
$r\approx\lsbar$.
\begin{figure}
  \psfrag{Gl3/k}{\small$Gl^3/\kappa$}\psfrag{lp/l}{\small$l_p/l$}
  \psfrag{lp1}{\small$l_p^1$}
\begin{center}
  \includegraphics[width=0.9\columnwidth]{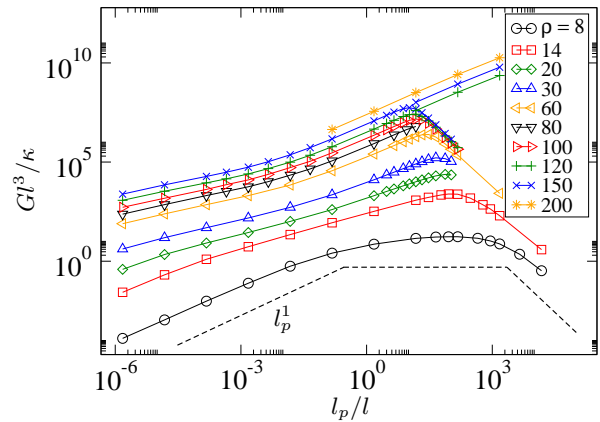}
\end{center}
\caption{Shear modulus $Gl^3/\kappa$ as function of persistence 
  length $l_p/l$ for various densities $\rho l$. The second branch in
  the upper right corner ($\rho l\ge120$) is obtained by suppressing
  the mechanical response (``$k_s\to\infty$''). The dashed line
  indicates the three regimes as obtained by Eq.~(\ref{eq:modFoam}).}
    \label{fig:modulus}
\end{figure}
This behavior is indicated by the dashed line in
Fig.~\ref{fig:modulus}, where it can be compared with the actual
results of our numerical analysis.
The normalized shear modulus $Gl^3/\kappa$ is shown as a function of
dimensionless persistence length $l_p/l$ for a set of dimensionless
densities $\rho l$.  At large $l_p/l$ (in the right part of the plot)
we recover purely mechanical behavior characterized by
$G\propto\bar{k}_s$ consistent with the mean-field picture of
Eq.~(\ref{eq:modFoam}) \cite{wil03,hea03a,ast00}.  Our main interest,
however, lies in the regime of small $l_p/l$ (in the left part of the
plot), where the persistence length is small enough for thermal
fluctuations to become relevant.  To analyze the modulus in the
thermal regime (where $k_s$ is irrelevant) it will be helpful to use
dimensional analysis and write the modulus in terms of the two
remaining response coefficients $\bar{k}_\perp$ and
$\bar{k}_\parallel$ of an average segment
\begin{equation}\label{eq:GDim}
  G(\kappa, l, l_p, \rho) = 
   \bar{k}_\perp g(\rho l,\bar{k}_\parallel/\bar{k}_\perp)\,.
\end{equation}
The first argument of the scaling function $g$, the density $x=\rho
l$, is of geometrical origin and counts the number of crosslinks per
filament.  The second argument,
$y=\bar{k}_\parallel/\bar{k}_\perp\simeq\rho l_p$, relates to the
energy balance between stretching and bending of an average segment
and marks a crossover at $y\approx1$ or $l_p\approx\lsbar$.  From
Fig.~\ref{fig:modulus} and the inset of Fig.~\ref{fig:modulusScaled}
one infers that for \emph{low densities} $g = y f(x)$, implying for
the modulus $G=\bar{k}_\parallel f(\rho l)$.  This linear dependence
on the ``pre-averaged'' stretching compliance $\bar{k}_\parallel$
hints at a foam-like stretching dominated regime~\cite{hea03c} where
correlations are absent. As one can also infer from these figures the
domain of validity of this linear regime is extremely narrow and
confined to low densities $x \leq 20$ and persistence lengths $y \ll
1$.
For \emph{medium and high densities} Fig.~\ref{fig:modulusScaled}
shows two non-trivial scaling regimes where $g(x\gg 1,y)= y^z$ becomes
independent of $x$ (and therefore of the filament length $l$) and
exhibits power law behavior with exponents $z=0.46$ and $z=0.9$ for
small and large values of $y$, respectively.  In both cases, the
modulus can be written in the form of a generalized geometric average
\begin{equation}\label{eq:GPoly} 
G \propto \bar{k}_\perp^{1-z}\bar{k}_\parallel^z\,,
\end{equation}  
which has to be contrasted with Eq.~(\ref{eq:modFoam}), where bending
and stretching modes are assumed to superimpose linearly.  Here,
correlation effects between the segments induce the non-trivial form
of the modulus and distinguish our fiber network from the ordinary
foam-like behavior obtained by single segment considerations. Whereas
foams may be considered as a limit where the filament length is zero
(or rather identical to the cell size) the scaling limit of fiber
networks corresponds to infinite fiber length.

To understand the origin of the correlations one has to take into
account the full distribution of segment lengths,
Eq.~(\ref{eq:segDist}). This will have a pronounced effect on an
affine deformation field $\delta_{\rm aff}\propto\gamma l_s$ as can be
seen by considering the axial force $f_\parallel$ along an affinely
stretched segment of length $l_s$, $f_\parallel =
k_\parallel\delta_{\rm aff} \simeq \kappa l_p^{\alpha}\gamma /
l_s^{2+\alpha}$.  In any but the purely mechanical situation, where
$\alpha=-2$ (and thus $f_s\simeq\kappa\gamma/r^2$), $f_\parallel$
strongly depends on the segment length.  This implies that, in
general, two neighboring segments on the same filament produce a net
force at their common node that has to be taken up by the crossing
filament which then, preferentially, will start to bend.  {F}rom the
exponential distribution of segment lengths in Eq.~(\ref{eq:segDist})
one can easily show that the size of these residual forces $\delta\!f$
can be arbitrarily large.  The corresponding probability distribution
$Q(\delta\!f)$ shows polynomial (fat) tails
\begin{equation}
Q(\delta\!f)\propto \delta\!f^{-4/3}P(l_s)\,, \delta\!f\to\infty
\end{equation}
and has a diverging mean value since $P(l_s=0)\neq 0$. As a
consequence there are always residual forces high enough to cause
significant bending of the crossing filament. Hence we conclude that
an affine deformation field is unstable and that the system can easily
lower its energy by redistributing the stresses to relieve shorter
segments and remove the tails of the residual force distribution
$Q(\delta\!f)$.

\begin{figure}
  \psfrag{ls/l}{\small$l_s/l$}\psfrag{E(ls)/Etot}{\small$E(l_s)/E_{tot}$}
\begin{center}
  \includegraphics[width=0.9\columnwidth]{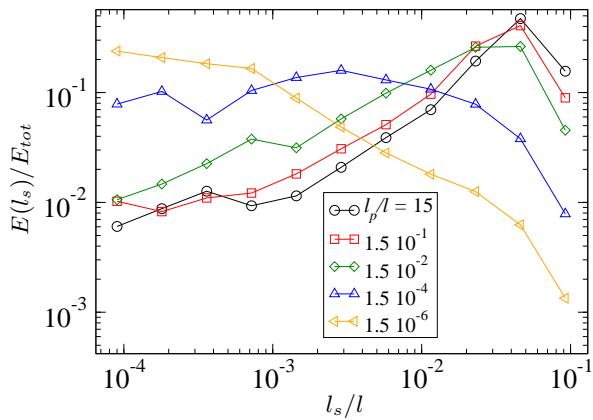}
\end{center}
\caption{Fraction of energy stored in the various segment lengths;
  the curves correspond to different persistence lengths at a density
  of $\rho l=80$, equivalent to $\lsbar/l\approx 2\cdot10^{-2}$.}
  \label{fig:enRatio}
\end{figure}

This mechanism can be used to derive an expression for the modulus in
the parameter region $y\ll1$, where the value of the exponent $z=0.46$
indicates that bending and stretching deformations contribute equally
to the elastic energy. We assume that segments up to a critical length
$l_c$ -- to be determined self-consistently -- will fully relax from
their affine reference state to give all their energy to the
neighboring segment on the crossing filament.  The energy of segments
with $l_s>l_c$ will then have two contributions. First, a stretching
part from the imposed affine strain field (for simplicity, we will set
$\alpha=1$ in what follows)
\begin{equation}\label{eq:strEnAff}
  w_s(l_s) \simeq k_\parallel\delta_{\rm aff}^2 \simeq 
\kappa\gamma^2\frac{l_p}{l_s^2}\,.
\end{equation}
Second, a bending part 
\begin{equation}\label{eq:bendEn}
  w_b(l_s) \simeq  k_\perp\delta_{\rm aff}^{'2} \simeq
  \kappa\gamma^2\frac{l^{'2}_s}{l_s^3} \, ,
\end{equation}
that only arises if the segment under consideration is neighbor to an
element on the crossing filament with $l'_s<l_c$ (the prime refers to
the neighboring small segment). Adding both contributions and
averaging over all segments $l_s>l_c$ and $l'_s<l_c$ we arrive at the
expression $w\simeq \kappa\gamma^2\rho(\rho l_p/x_c+x_c)$, where $x_c
:= \rho l_c \ll 1$ in the parameter range of interest.  Minimizing
with respect to $x_c$ gives the required expressions $x_c^{\rm min}
\simeq (l_p\rho)^{1/2}$ and $G\simeq\rho^2 w_{\rm
  min}/\gamma^2\simeq\kappa\rho^{7/2}l_p^{1/2}$ corresponding to a
value $z=1/2$ for the exponent that compares well with the measured
value $z=0.46$.  Repeating the calculation for general values of
$\alpha$ gives $z(\alpha)=\alpha/(1+\alpha)$. We have verified this
result by simulations with an accuracy of about ten per
cent~\cite{tbp}. The non-trivial behavior of $G$ observed in
Figs.~\ref{fig:modulus} and \ref{fig:modulusScaled} can thus be
explained by a length scale $l_c=\lsbar(l_p/\lsbar)^{1/2}$ below which
the affinity of the deformation field breaks down.  The mechanism is
illustrated in Fig.~\ref{fig:enRatio}, where a histogram for the
fraction of energy stored in segments of various lengths is shown.
Increasing the persistence length, the short segments one after the
other loose their energies in favor of additional excitations in
longer segments.

When, eventually, $l_c\approx l_p\approx\lsbar$ ($y\approx1$) the
affine strain field does not serve as a reference configuration any
more, since it is strongly perturbed by a majority of segments with
$l_s<l_c$.  Moreover, the unloading of the smaller segments now
produces significant stretching deformations of their neighbors on the
{\em same} filament such that the available energy for bending of the
{\em crossing} filament is reduced.  At this stage, one enters the
second intermediate asymptotic regime where, like in the affine regime
at low densities, stretching modes dominate the modulus. As can be
seen in Fig.~\ref{fig:enRatio}, only the longest segments carry
substantial amounts of energy such that the displacement field must be
highly non-affine. For $l_p/l\ge 1.5\;10^{-2}$ about $90\%$ of the
energy is stored in the longest $30\%$ of the segments. In this
parameter range, bending is on average the softer mode
$y=\bar{k}_\parallel/\bar{k}_\perp\gg1$ and therefore contributes only
very little to the total energy. Raising the density to still higher
values, it is conceivable from our data that the exponent $z=0.9$
approaches $z=1$, which would mean that the energy in the bending
modes is completely negligible and $G\sim \bar{k}_\parallel$ as in the
affine regime.  A transition into a regime dominated by the low-energy
bending modes would not be favorable, however. As is known from the
mechanical fiber model~\cite{wil03}, such a regime must not be
characterized by the pre-averaged force constant
$\bar{k}_\perp=k_\perp(\lsbar)$ but by an effective stiffness $\langle
k_\perp\rangle \simeq \kappa/\xi^3$ with a new length scale $\xi=
l(\rho l)^{-\mu/3}$ and $\mu\approx6.7$ that is highly dependent on
fiber length $l$.

In summary, we found that for a broad range of parameters the
macroscopic shear modulus is asymptotically independent of the fiber
length. Affine stretching is energetically unstable towards a
redistribution of energies in favor of longer segments. This gives
rise to a correlation-induced elasticity that cannot be explained
within a ``single cell'' picture. In addition to a non-trivial mixed
regime we found a stretching dominated regime that is characterized by
exponents similar to those obtained in the foam-like stretching
regime.  Microscopically, however, the stress is distributed in a
completely different way and is dominated by very few, very long fiber
segments.  As mentioned above, one particular application of our model
is to the macroscopic linear response of stiff polymer networks. The
results may therefore be directly relevant for two-dimensional
networks of the filamentous biopolymer F-actin, assembled on top of
micro-fabricated pillars \cite{roo03}. In addition, it might shed new
light on very recent rheological measurements on cross-linked actin
networks \cite{gar04b,shi04}, which emphasize the single-polymer
origin of the measured elastic moduli. Our simulations, on the
contrary, highlight the potentially non-trivial effects of
inter-polymer correlations on the macroscopic elasticity.


\end{document}